# Development of SNSPD System with Gifford-McMahon Cryocooler

Shigehito Miki, Mikio Fujiwara, Masahide Sasaki, and Zhen Wang

*Abstract*— A superconducting nanowire single photon detector (SNSPD) system for telecommunication wavelength using a GM cryocooler was developed and its performance was verified. The cryocooler based SNSPD system can operate continuously with a 100 V AC power supply without any cryogen. The packaged SNSPD device was cooled to 2.96 K within a thermal fluctuation range of 10 mK. An SNSPD with an area of 20 x 20 $\mu m^2$ showed good system detection efficiency (DE) at 100 Hz dark count rate of 2.6% and 4.5% at wavelengths of 1550 and 1310 nm, respectively. An SNSPD with an area of 10 x 10 $\mu m^2$ and kinetic inductance lower than that of the large area device showed good system DE of 2.6% at a wavelength of 1550 nm. The SNSPD system could be operated for over 10 h with constant system DE and dark count rate.

## I. INTRODUCTION

A practically available single photon detector for operation in the telecommunication wavelength (1550 nm) is essential for quantum information processing and sensing. Although semiconductor avalanche photo diodes (APD) have been commonly used thus far for quantum optical experiments, their performances are not satisfactory and they require complicated operations such as wavelength conversion and precise gating to avoid an after pulse. In contrast, a superconducting nanowire single photon detector (SNSPD) has potentially single photon sensitivity from the visible to near infrared wavelength range, excellent timing resolution, high counting rate up to several GHz, and low dark count rate[1-3]. In addition, the SNSPD can be operated without requiring a wavelength converter and gating system, making it more attractive than the APD system.

Although research on the SNSPD is still at a nascent stage, practical SNSPD systems realized with the current level of achievable performance are still attractive for quantum key distribution (QKD) and other quantum information processing applications because of their superiority over APD devices. In fact, such systems have been successfully employed in QKD experiments [4,5], and they have been shown to boost both the transmission distance and the key generation rate. For practical use, an SNSPD system should be compact, facilitate easy and stable operation, and provide high performance. An SNSPD system based on a small Gifford-McMahon (GM) cryocooler is an adequate choice in this regard because it can operate continuously without any cryogen whilst requiring low input power [6,7]. To apply a GM cryocooler based SNSPD system to quantum processing applications, it is necessary to ensure adequate specifications of the system and verify its performance.

Needless to say, further improvements are highly desirable, and significant effort is being made to increase the detection efficiency (DE) and continuous photon-counting rate of these systems. The DE is mainly determined by the intrinsic photon absorption coefficient, effective optical coupling coefficient to the active area, and device detection efficiency. It is necessary to improve all limiting factors in order to obtain high system detection efficiency. The primary factor limiting the counting rate is the large kinetic inductance (KI) of the nanowire[8,9]; this factor determines the device recovery time after photon absorption. Reducing the active area is a natural way to reduce the KI, however this makes efficient optical coupling more challenging.

In this paper, we report on the development of an SNSPD system and verify its performance from the practical point of view by considering system detection efficiency at nominal dark count rate, system continuous counting rate, and system operation stability versus running time. We also describe several results indicating improvement in SNSPD system performance.

## II. DEVELOPMENT OF SNSPD SYSTEM

For developing an SNSPD system, we used a two-stage type small Gifford-McMahon (GM) cryocooler. The rated input power consumption was 1.5 kW at a driving frequency of 60 Hz. The sample stage for cooling SNSPD packages was connected to the second stage through a stainless steel plate and a lead block with large heat capacity to reduce thermal fluctuation[6]. The sample stage could be cooled to 2.96 K within a thermal fluctuation range of 10 mK. To achieve efficient coupling

Manuscript received August 15, 2008.

Shigehito Miki and Zhen Wang are with the Kobe Advanced Research Center, National Institute of Information and Communications Technology, 588-2, Iwaoka, Nishi-ku, Kobe, Hyogo 651-2492, Japan (e-mail: s-miki@nict.go.jp).

Mikio Fujiwara and Masahide Sasaki are with the National Institute of Information and Communications Technology, 4-2-1 Nukui-Kitamachi, Koganei, Tokyo 184-8795, Japan.

TABLE.1 NANOWIRE DESIGN, NANOWIRE DC CHARACTERISTICS, AND MEASURED DEVICE PERFORMANCE FOR TWO (PLUS ONE REFERRED IN [9]) TYPICAL SNSPD DESIGNS.

|  | Ref [9] | #A | #B |
|---|---|---|---|
| Device size (μm) | 20 x 20 | 20 x 20 | 10 x 10 |
| Line width (nm) | 100 | 100 | 100 |
| Line pitch (nm) | 200 | 200 | 200 |
| films thickness (nm) | 4.2 | <4 | <4 |
| $T_c$ (K) | 12 | 9.8 | 9.5 |
| $I_c$ (μA) | 32 | 19 | 15 |
| $R_{20K}$ (MΩ) | 7 | ~10 | ~2.5 |
| System DE (%) | 0.4-3.5 at 2.5K | 0.8-2.6 at 2.95 K | 2.6 at 2.95 K |
| $L_{k,device}$ (μH) | 1.0 | 2.0 | 0.44 |

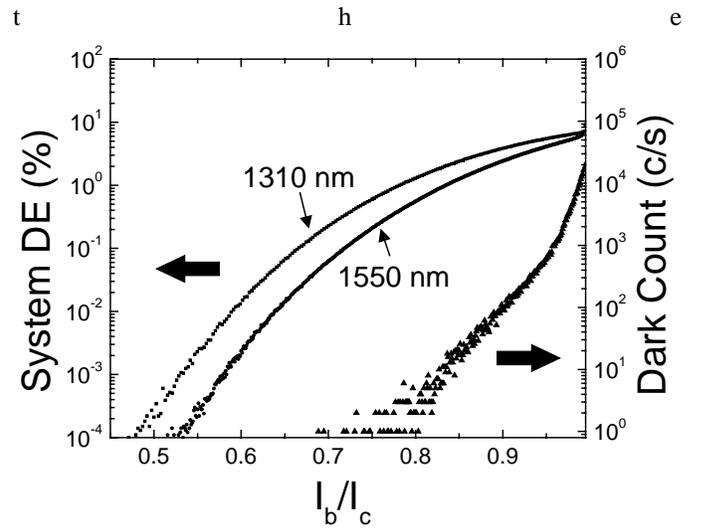

Figure 1. System DE and dark count rate as a function of bias current normalized by Ic for a SNSPD device with design #A.

between the incident photons and the meander area, we prepared SNSPD packages consisting of a pair of oxygen-free copper blocks: one for mounting the device and the other to fix the end of the single mode optical fiber with a core diameter of 9 μm. Prior to cooling, the blocks were accurately aligned so that the light spot was efficiently coupled to the meander area by monitoring the device and light spot from the backside of the substrate. After careful adjustment, the SNSPD packages were set on the sample stage. Up to six SNSPD packages could be set in a cryocooler, and we introduced brass semi-rigid coaxial cables for each package to bias the current and we read the output signals. A 1550 (1310) nm wavelength continuous laser diode was used as the input photon source, and it was heavily attenuated so that the photon flux at the input connector of the cryostat was $10^7$ photons/s. A fiber polarization controller was inserted in front of the cryocooler optical input to control the polarization properties of the incident photons so that their polarization sensitivity (maximizing the DE) matched that of each device. The output port was connected to a bias tee and two low noise amplifiers (LNAs) through a coaxial cable at room temperature. The device was current biased via the dc arm of the bias tee, and the output signal was counted through the ac arm of the bias tee and two LNAs. The system detection efficiency was defined as output count rate divided by the photon flux rate input to the system. The superconducting NbN nanowire devices were fabricated on single crystal MgO substrates. A method for growing ultrathin NbN epitaxial films and the process for fabricating nanowire devices are already described elsewhere [10].

## III. PERFORMANCE EVALUATION

The nanowire geometry design, dc characteristics, and device performance measured by our SNSPD system for two (plus one referred in [11]) typical device designs are listed in Table 1. Since the thickness of the nanowire was less than 4 nm, devices showed lower $T_c$ and smaller $I_c$ as compared to those described in [11]. Figure 1 shows the system DE and dark count rate as a function of the bias current normalized by $I_c$ for an SNSPD device with design #A. The system DEs (at 100 Hz dark count rate) of the device were about 2.6% and 4.5% at wavelengths of 1550 and 1310 nm, respectively. These values were almost the same as those obtained in our previous results taking into account the higher operation temperature. The intrinsic DE of the NbN nanowires seemed higher according to the results of [12]. It is difficult to distinguish among the possible causes limiting the system DE at this stage. However, it is necessary to consider the various factors sequentially to improve the system DE further, such as device design (nanowire width, size, thickness, and filling factor), extent of constrictions, and operating temperature. For example, we do not see a saturation in the detection efficiency at high bias as observed elsewhere [12], indicating the device detection efficiency is insufficient due to unoptimized nanowire design, constriction in a nanowire, and higher operation temperature. Further optimization of nanowire design and reduction of constrictions will facilitate the improvement of device detection efficiency.

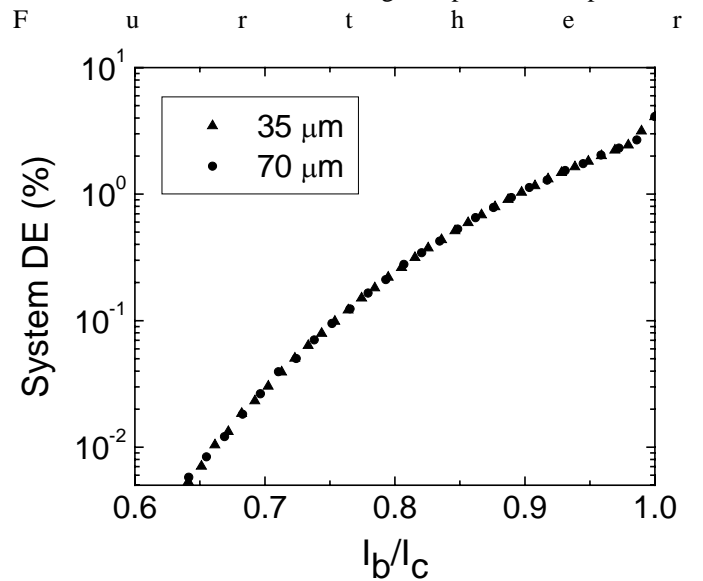

Figure 2. System DE of device with design #A as a function of the bias current normalized by $I_c$ at different fiber end-device distances of 35 μm and 70 μm.

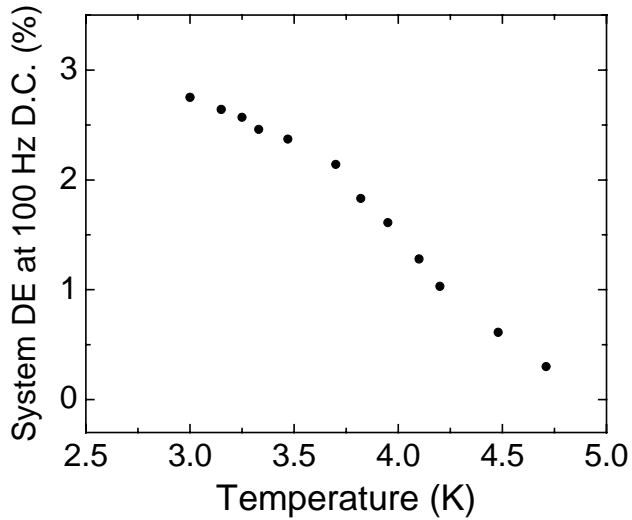

Figure 3. System DE at 100 Hz dark count rate of device with design #A as a function of the operation temperature.

Figure 2 shows the system DE as a function of the operation temperature. The bias current was set to correspond to a dark count rate of 100 Hz and the system DE was measured at each temperature. The dark count reduced with the temperature, enabling the biasing of the current closer to Ic. Therefore, the system DE increased with a decrease in the temperature. The lowest temperature achieved in our system was even higher that those in other reports. Therefore, further improvement in the system DE might be expected if a lower operating temperature is achieved in future.

Since the optical coupling efficiency is also important factor limiting the system DE, we measured the system DE of the same device twice by changing the distance between the end of the fiber and the meander to 35 μm and 70 μm, as shown in Fig. 3. The distances were assessed from swept-frequency laser interferometry[13]. The results showed that the system DE did not change, indicating that a 20 x 20 μm$^2$ device was large enough to couple the light spot to the meander area and there was no misalignment after cooling. It is natural to assume that a smaller device with a smaller kinetic inductance could be coupled to efficiently if the distance between the end of fiber and the device were reduced. Therefore, we mounted the 10 x 10 μm$^2$ nanowire devices and set the distance between the end of the fiber and the device to 20 μm before cooling. The results showed that the 10 x 10 μm$^2$ area devices can achieve a system DE of 2.6 %. In addition, the device kinetic inductance, measured by observing the phase of a reflected microwave signal versus the frequency using a network analyzer [11,12], can be reduced from 2.0 μH to 0.44 μH.

We next investigated the stability of our SNSPD system by running it for over 10 h. Figure 4 shows the system DE, dark count, and device temperature as a function of the running time. The system was always biased at the same direct current to achieve a dark count rate of 100 Hz. As shown in the figure, our SNSPD system was able to operate with a constant system DE of 2.6% and dark count rate of 100 Hz. It is supposed that our SNSPD system can operate continuously with stable performance. Although a sudden increase in the dark count caused by disturbance from the external environment was observed on a rare occasion, the increase disappeared after 8 h (PM 8:00–10:00) at which time the noise environment was good. Finally, figure 5 shows the histogram of the measured #A device system DE. We measured 12 devices in all; most of the devices exceed a system DE of 1%, enabling their application to quantum optical experiments that need multiple detectors.

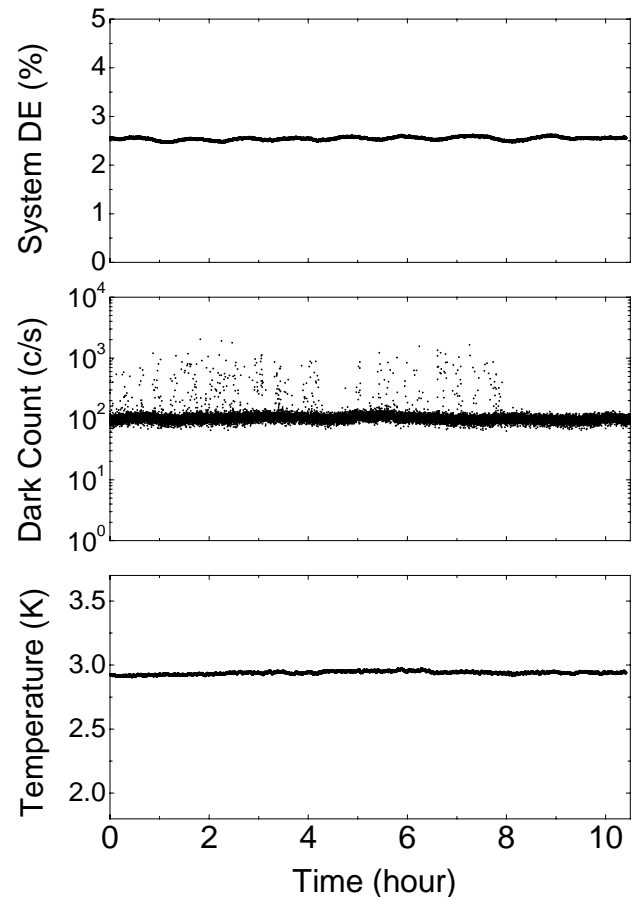

Figure 4. System DE, dark count rate, and operation temperature of SNSPD system as a function of the running time. Device with design #A was mounted into the system.

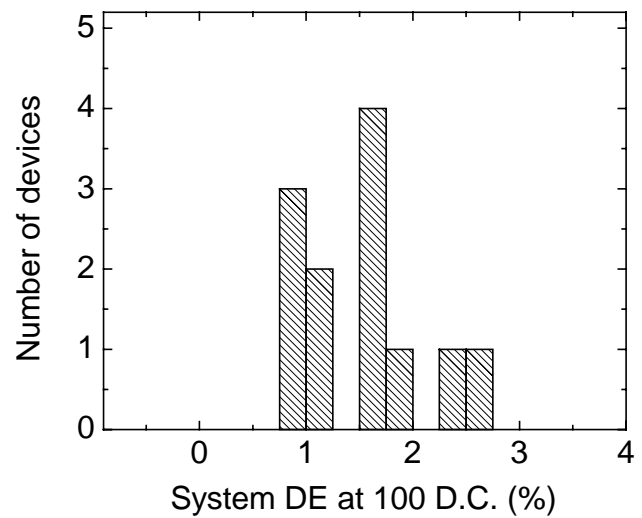

Figure 5. Histogram of system DE at 100 Hz dark count rate for 12 measured devices with design #A.

## IV. CONCLUSION

We developed a GM cryocooler based SNSPD system and evaluated its performance. By mounting large 20 x 20 $\mu m^2$ area nanowire devices, we found that over 10 devices demonstrated relatively higher system DEs, greater than 1%, at a wavelength of 1550 nm. Small 10 x 10 $\mu m^2$ area devices also showed a system DE of 2.6% and low kinetic inductance of 0.44 $\mu H$, implying minimum continuous count rate of 63 MHz could be achieved with our system. We verified that our SNSPD system can operate continuously for over 10 h with constant DE and dark count rate performance. These results indicate that our developed system will enable improved performance in QKD and other photon counting applications, and further improvement of our SNSPD system performance will broaden the impact of SNSPDs in these applications.

## V. ACKNOWLEDGEMENT

We would like to thank Sae Woo Nam, Burm Baek at National Institute of Standards and Technology, Aaron J. Miller at Albion College, and Robert H. Hadfield at Heriot-Watt University for helpful advise and discussions to develop the SNSPD system.

## REFERENCES

[1] G. Gol'tsman, O. Okunev, G. Chulkova, A. Lipatov, A. Semenov, K. Smirnov, B. Voronov, A. Dzardanov, C. Williams, and R. Sobolewski, "Picosecond superconducting single photon detector," Appl. Phys. Lett., vol. 79, 2001, pp. 705–707.
[2] A. Korneev, V. Matvienko, O. Minaeva, I. Milostnaya, I. Rubtsova, G. Chulkova, K. Smirnov, V. Voronov, G. Gol'tsman, W. Slysz, A. Pearlman, A. Verevkin, and R. Sobolewski, "Quantum Efficiency and Noise Equivalent Power of Nanostructured, NbN, Single-Photon Detectors in the Wavelength from the Visible to Infrared," IEEE. Trans. Appl. Supercond., vol. 15, 2005, pp. 571–574.
[3] K. M. Rosfjord, J. K. W. Yang, E. A. Dauler, A. J. Kerman, V. Anant, B. M. Boronov, G. N. Gol'tsman, and K. K. Berggren, "Nanowire single photon detector with an integrated optical cavity and anti-reflection coating," Optics Express, vol. 14, 2006, pp. 527-534.
[4] R. H. Hadfield, J. L. Habif, J. Schlafer, R. E. Schwall, " Quantum key distribution at 1550 nm with twin superconducting single-photon detectors," Appl. Phys. Lett., vol. 89, 2006, pp. 241129.
[5] H. Takesue, S. Nam, Q. Zhang, R. H. Hadfield, Y. Yamamoto, " Quantum key distribution over a 40-dB channel loss using superconducting single-photon detectors," Nat. Photonics, vol.1, 2007, pp.343-348.
[6] R. H. Hadfield, M. J. Stevens, S. S. Gruber, A. J. Miller, R. E. Schwall, R. P. Mirin, S. W. Nam, "Single photon source characterization with a superconducting single photon detector," Opt. Express, vol.13, no.26, 2005, pp. 10847-10853
[7] R. Radebaugh "Refrigeration for Superconductors," Proceedings of the IEEE, vol. 92, 2004, pp.1719-1734.
[8] R. H. Hadfield, A. J. Miller, S. W. Nam, R. L. Kautz, R. E. Schwall, " Low-frequency phase locking in high-inductance superconducting nanowires," Appl. Phys. Lett., vol. 87, 2005, pp.203505.
[9] A. J. Kerman, E. A. Dauler, W. E. Keicher, J. K. W. Yang, K. K. Berggren, G. Gol'tsman, B. Voronov, " Kinetic-inductance-limited reset time of superconducting nanowire photon counters," Appl. Phys. Lett., vol. 88, 2006, pp. 111116.
[10] S. Miki, M. Fujiwara, M. Sasaki, and Z. Wang, " NbN Superconducting Single-photon Detectors Prepared on Single-crystal MgO Substrates," IEEE Trans. Appl. Supercond., vol. 17, 2007, pp.285-288.
[11] S. Miki, M. Fujiwara, M. Sasaki, B. Baek, A. J. Miller, R. H. Hadfield, S. W. Nam, and Z. Wang, " Large sensitive-area NbN nanowire superconducting single-photon detectors fabricated on single-crystal MgO substrates," Appl. Phys. Lett., vol. 92, 2008, pp.061116.
[12] A. J. Kerman, E. A. Dauler, J. K. W. Yang, K. M. Rosfjord, V. Anant, K. Berggren, G. N. Gol'tsman, and B. M. Voronov, " Constriction-limited detection efficiency of superconducting nanowire single-photon detectors," Appl. Phys. Lett., vol. 90, 2007, pp.101110.
[13] A. E. Lita, A. J. Miller, S. W. Nam, "Counting near-infrared single-photons with 95% efficiency," Optics Express, vol. 16, 2008, pp. 3032-3040.